\documentclass[prb,twocolumn,showpacs,amsmath,amssymb,superscriptaddress]{revtex4-1}
\usepackage{graphicx}
\usepackage{breqn}
\usepackage{xcolor}
\usepackage{color}
\usepackage{booktabs}
\usepackage{float}
\usepackage{longtable}
\usepackage{epsfig}
\usepackage{bm}
\usepackage{dcolumn}
\usepackage{lipsum, babel}
\usepackage{soul}
\usepackage{amssymb}% http://ctan.org/pkg/amssymb
\usepackage{pifont}% http://ctan.org/pkg/pifont
\newcommand{\cmark}{\ding{51}}%
\newcommand{\xmark}{\ding{55}}%

\usepackage{rotating} % for sidewaystable

\usepackage[normalem]{ulem}
\usepackage{epstopdf}
\makeatletter
\let\cat@comma@active\@empty
\makeatother
\begin{document}

%\title{Altermagnets:\\   Superfluid $^3$He-like phases that may well be not the most useless}
%\title{Robust realization of spin-ordered anisotropic zero-magnetization phase in altermagnets}
%\title{Altermagnetism versus superfluid $^3$He and Fermi-liquid instabilities:\\ Analogies and distinctions}
%\title{Altermagnetism: \\ Analogies and distinctions from superfluid $^3$He and Fermi-liquid instabilities}
\title{Altermagnetism: an unconventional spin-ordered phase of matter}
\author{T.~Jungwirth}
\affiliation{Institute of Physics, Czech Academy of Sciences, Cukrovarnick\'a 10, 162 00, Praha 6, Czech Republic}
\affiliation{School of Physics and Astronomy, University of Nottingham, NG7 2RD, Nottingham, United Kingdom}
\author{R. M. Fernandes}
\affiliation{Department of Physics, University of Illinois Urbana-Champaign, Urbana, IL 61801, USA}
\affiliation{Anthony J. Leggett Institute for Condensed Matter Theory, University of Illinois at Urbana-Champaign,
IL 61801, USA}
\author{E. Fradkin}
\affiliation{Department of Physics, University of Illinois Urbana-Champaign, Urbana, IL 61801, USA}
\affiliation{Anthony J. Leggett Institute for Condensed Matter Theory, University of Illinois at Urbana-Champaign,
IL 61801, USA}
\author{A. H. MacDonald}
\affiliation{Department of Physics, The University of Texas at Austin, Austin, TX 78712, USA}
\author{J. Sinova}
\affiliation{Institut f\"ur Physik, Johannes Gutenberg Universit\"at Mainz, D-55099 Mainz, Germany}
\author{L. \v{S}mejkal}
\affiliation{Max Planck Institute for the Physics of Complex Systems, N\"othnitzer Str. 38, 01187 Dresden, Germany}
\affiliation{Max Planck Institute for Chemical Physics of Solids, N\"othnitzer Str. 40, 01187, Dresden, Germany}
\affiliation{Institute of Physics, Czech Academy of Sciences, Cukrovarnick\'a 10, 162 00, Praha 6, Czech Republic}

\date{\today}

\begin{abstract}
The Pauli exclusion principle combined with interactions between fermions is a basic mechanism across condensed-matter systems giving rise to a spontaneous breaking of the spin-space rotation symmetry of spin-ordered phases. Ferromagnetism is a conventional manifestation of spin ordering which leads to numerous applications, e.g., in spintronic information technologies. Altermagnetism, whose recent discovery was largely motivated by spintronics, stands apart from conventional magnetism in the sense that it spontaneously breaks not only spin-space but also real-space rotation symmetries, while it preserves a symmetry combining spin-space and real-space rotations. This is realized on crystals by a collinear compensated ordering of spins with a characteristic $d$, $g$ or $i$-wave symmetry.  Our Perspective goes beyond the theory of spin arrangements on crystals by connecting altermagnetism to basic notions in condensed matter physics. Specifically, we reflect on the analogies and distinctions of altermagnetism as compared to superfluid $^3$He and theories of spin ordering in the momentum space generated by other higher-partial-wave instabilities of a Fermi-liquid. On one hand, all these physical systems have in common the extraordinary combination of spontaneous breaking of spin-space and real-space rotation symmetries. On the other hand, we point out that there are key differences, both at the symmetry level and, particularly, at the level of microscopic mechanisms of ordering. These explain the comparatively large abundance, robustness  and utility of altermagnetism, as predicted by the symmetry-classification of spin arrangements on crystals and {\em ab initio} calculations, and supported by  initial experiments.
\end{abstract}

\maketitle

% Word count without equations and figure captions: 2835
% figures: 5x aspsect ratio 1 gives: 5 x 170=850
% equations: 2x32=64
% captions: 384
% total: 4133
% PRL total limit 3750
%\subsection{Introduction}
%\label{intro}

%**************************************************************************
\subsection{Overview}
\label{overview}

\begin{figure}[h!]
	\centering
	\includegraphics[width=1\linewidth]{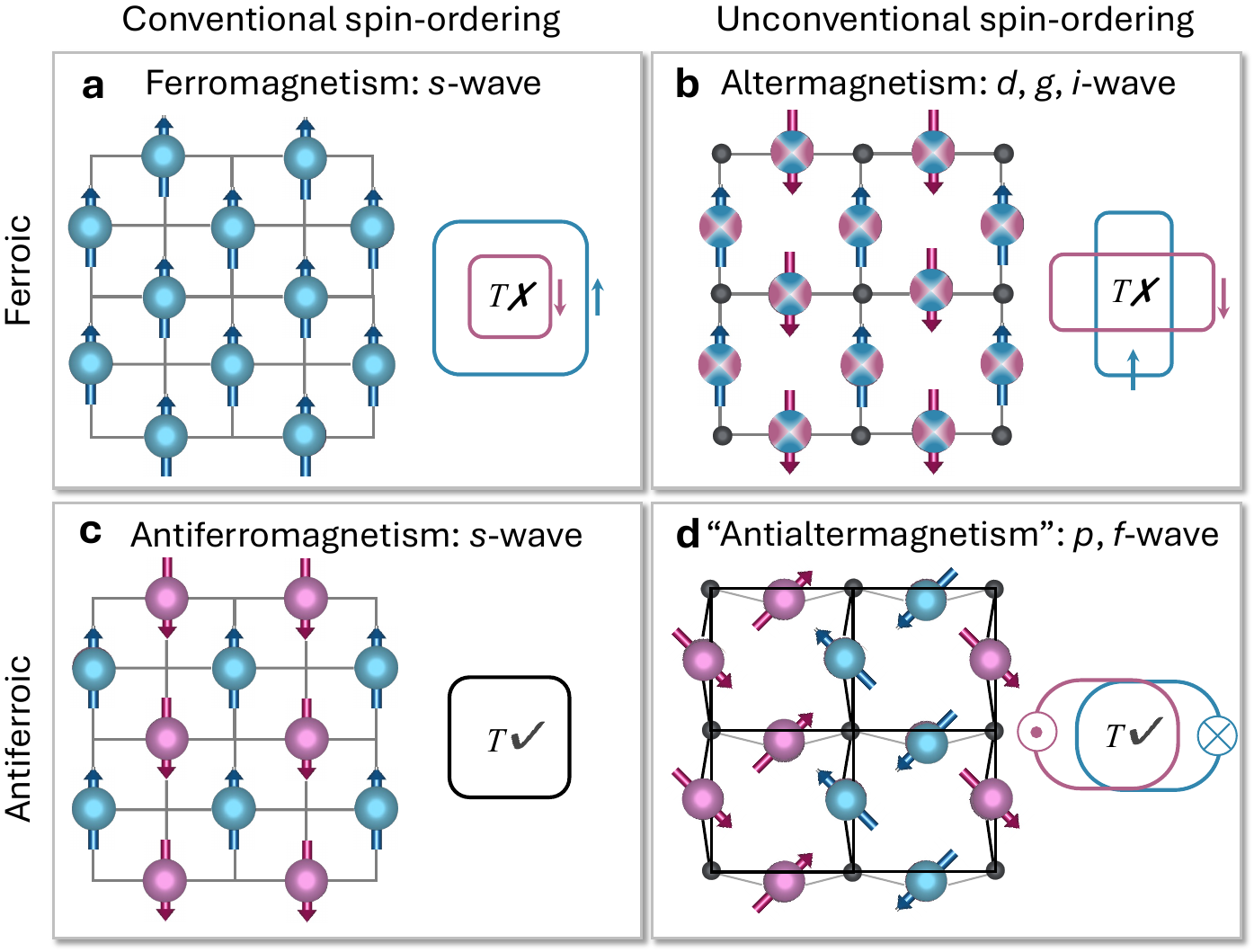}
	\caption{
	\textbf{Cartoons of the correspondence between the  spin arrangements on  crystals and the momentum-space spin-dependent energy iso-surfaces.}
(a) Conventional collinear ferromagnetism  with a ferroic order of local atomic dipoles in the position space  and corresponding majority-spin (blue) and minority-spin (red) energy iso-surfaces in the momentum space preserving the crystallographic point-group rotation symmetry ($s$-wave). (b)  Collinear altermagnetism with depicted superimposed dipole component (blue or red arrows) and a higher-partial-wave, specifically $d$-wave component (blue-red spheres) of  the local anisotropic spin density on atomic sites. The higher-partial-wave  ($d$-wave) component is ferroically ordered on the crystal. Blue and red colors mark opposite spin polarizations.   The momentum-space spin-up and spin-down iso-surfaces show the corresponding unconventional higher-partial-wave ($d$-wave) spin ordering, breaking the crystallographic point-group rotation symmetry. The ferroic order  leads to broken time-reversal symmetry ({\em T}\,\xmark) in both ferromagnetism an altermagnetism. (d) ``Antialtermagnetism" has a non-collinear coplanar spin arrangement on the crystal. It shares with conventional collinear N\'eel antiferromagnetism (c) an antiferroic order of local atomic dipoles, characterized by a symmetry combining time-reversal and translation. This is reflected in preserved time-reversal symmetry ({\em T}\,\cmark) in their spin-polarized (antialtermagnetism) and spin-degenerate (N\'eel antiferromagnetism) momentum-space electronic structures. The crystallographic point-group rotation symmetry is preserved in the momentum-space iso-surface of the conventional N\'eel antiferromagnetism, similar to the conventional ferromagnetism. In ``antialtermagnetism",  the  momentum-space electronic structure shares with altermagnetism the unconventional breaking of the crystallographic point-group rotation symmetry, as well as the collinear spin polarization. In ``antialtermagnetism", the collinear spin polarization in the momentum space is orthogonal to the coplanar spins in the position space of the crystal lattice.}
\label{fig_FM_AM_AFM_AAM}
\end{figure}

The research beyond conventional magnetism, which led to the recent delineation of the altermagnetic symmetry class\cite{Smejkal2021a}, was largely motivated by the field of spintronics\cite{Smejkal2022AHEReview,Smejkal2022a}. From an applied perspective, spintronics is a modern branch of integrated-circuit technologies currently undergoing a transition from niche to mass production, in particular thanks to embedded non-volatile memories complementing semiconductors on advance-node processor chips\cite{Lee2022a,Ambrosi2023,IRDS2023}. The functionality of present spintronic memories is based on the magnetization in conventional ferromagnets which generates well separated and conserved spin-up and spin-down channels in the electronic structure  (Fig.~\ref{fig_FM_AM_AFM_AAM}a). Simultaneously, however, the magnetization sets physical limits on the spatial, temporal and energy scalability of the spintronic technology\cite{Smejkal2022AHEReview,Smejkal2022a}. Altermagnetism opens a prospect of removing these limits by combining well separated and conserved spin-up and spin-down channels with vanishing net magnetization (Fig.~\ref{fig_FM_AM_AFM_AAM}b)\cite{Smejkal2021a,Smejkal2022a}. It enables this extraordinary combination of properties thanks to the unconventional anisotropic $d$, $g$, or $i$-wave nature of its magnetic ordering. 

In this article we dissect the altermagnetic ordering from the symmetry and microscopic-mechanism perspectives, and connect it to basic condensed-matter physics notions developed in the research of instabilities in metallic Fermi liquids, namely of superfluid $^3$He\cite{Leggett1975,Vollhardt1990,Wolfle1999,Leggett2003,Brison2021,Volovik2022} and higher-partial-wave spin-channel Pomeranchuk instabilities\cite{Akhiezer1978,Hirsch1990,Marchenko1991,Gorkov1992,Oganesyan2001,Kivelson2003,Wu2004,Wu2007,Chubukov2009,Alexandradinata2010,Fischer2011,Kiselev2017,Wu2018,Klein2019}. Apart from analogies, we highlight key distinctions of altermagnetism to reflect on its comparatively large abundance, robustness, and foreseen utility in diverse science and technology fields. Indeed, altermagnetism was predicted to emerge in numerous materials, covering a broad range of interaction strengths and conduction types from  weakly-interacting metals to strongly-correlated Mott insulators\cite{Smejkal2020,Smejkal2021a,Smejkal2022a,,Guo2023b,Xiao2023,Jungwirth2024}, and to enable interplay  with semiconducting, superconducting or ferroelectric phases\cite{Smejkal2021a,Smejkal2022a,Betancourt2021,Mazin2022,Sun2023,Papaj2023,Beenakker2023,Zhu2023d,Gu2024,Sheng2024,Smejkal2024}. Experimentally, altermagnetism has been already observed in materials  with  ordering above room temperature\cite{Krempasky2024,Lee2024,Osumi2024,Hajlaoui2024,Chilcote2024,Amin2024,Reimers2024,Yang2024,Ding2024,Zeng2024,Li2024,Lu2024}. 

As with any spin-ordered phase, altermagnetism shares with superfluid $^3$He and with the ordered phases generated by the spin-channel Pomeranchuk instabilities the spontaneously broken continuous spin-space rotation symmetry\cite{Andreev1980,Andreev1984,Vollhardt1990,Gorkov1992,Wu2004,Moore2010a,Smejkal2021a,Moessner2021}. In addition, altermagnetism, superfluid $^3$He and the ordered phases generated by the higher-partial-wave spin-channel Pomeranchuk instabilities also spontaneously break the  real-space rotation symmetry, and by doing so they stand apart from  conventional magnetism\cite{Leggett1975,Vollhardt1990,Wolfle1999,Leggett2003,Brison2021,Smejkal2021a,Volovik2022,Akhiezer1978,Hirsch1990,Marchenko1991,Gorkov1992,Oganesyan2001,Kivelson2003,Wu2004,Wu2007}. 

In altermagnetism, the  real-space symmetries concern discrete crystallographically-constrained rotations (proper or improper and symmorphic or non-symmorphic). Remarkably, their spontaneous breaking in the ordered ground state can be already prearranged by the crystal structure in the normal phase, which underlines the inherent role of the crystal lattice in altermagnetism. In contrast, $^3$He is a uniform liquid of fermionic particles in a free space with no underlying lattice. In the normal phase, both the spin-space and the real-space rotation symmetries are thus continuous, and they spontaneously break, hand-in-hand, upon the transition to the ordered superfluid phase\cite{Leggett1975,Vollhardt1990,Wolfle1999,Leggett2003,Brison2021,Volovik2022}. Similarly, the theory of Pomeranchuk instabilities  commonly assumes a uniform Galilean-invariant Fermi liquid or, if included, the lattice effects enter only indirectly via non-Galilean anisotropic or non-quadratic corrections to the single-particle energy dispersion in the momentum space\cite{Hirsch1990,Wu2004,Wu2007}. 

\begin{figure}[h!]
	\centering
	\includegraphics[width=1\linewidth]{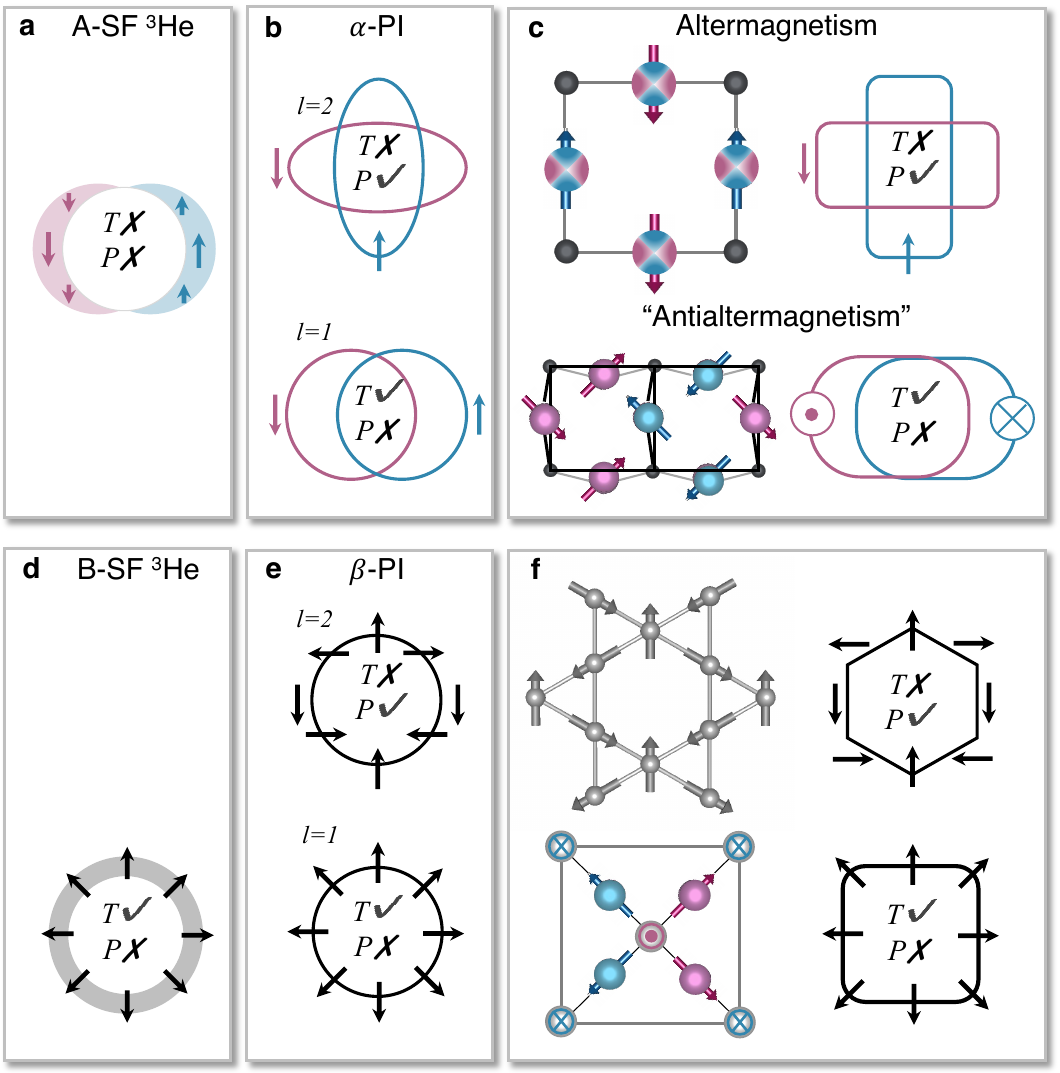}
	\caption{
	\textbf{Cartoons of ordered phases with spontaneous breaking of spin-space and real-space rotation symmetries.}
(a) The  A-phase of superfluid $^3$He with the anisotropic Cooper-pairing (gap) function in the momentum space. Arrows represent the order-parameter,  ${\bf d}^{\rm A}_{l=1}({\bf k})\propto (0, ik_x - k_y, 0)$, which is a vector in the spin space with one non-zero component given by the $l_z=1$ spherical harmonic expressed in the momentum space. (b) The $\alpha$-phase spin-channel $l=2$ and $l=1$  Pomeranchuk instabilities of a Fermi liquid with arrows showing the momentum-dependent collinear spin polarization on the anisotropically distorted/shifted Fermi surfaces.  (c) Representative altermagnetic and ``antialtermagnetic" crystals with analogous momentum-space collinear spin polarization and energy iso-surface distortion/shift to the $\alpha$-phase Pomeranchuk instabilities.  (d) The B-phase of superfluid $^3$He with the isotropic Cooper-pairing (gap) function in the momentum space. Arrows represent the order-parameter spin-space vector,  ${\bf d}^{\rm B}_{l=1}({\bf k})\propto{\bf k}$, whose components are given by a combination of  all three spherical harmonics $l_z=\pm1$ and 0. (e) The $\beta$-phase spin-channel $l=2$ and $l=1$  Pomeranchuk instabilities of a Fermi liquid with arrows showing the momentum-dependent spin texture on isotropic Fermi surfaces.  (For brevity we show only one of the two spin-split Fermi surfaces.) (f) Representative magnetic-crystals with analogous momentum-space spin textures to the $\beta$-phase Pomeranchuk instabilities. In all panels, broken/preserved time-reversal ($T$) and inversion (parity $P$) symmetries in the momentum space are shown by \xmark/\cmark marks. 
%CL/CP/NCP {\bf s}({\bf r}) refers to collinear, non-collinear coplanar, and non-coplanar spin arrangement in the position space of the crystal lattice. T t labels a symmetry combining time-reversal and translation, T R$_{\rm r}$ time-reversal and real-space rotation, and R$_{\rm s}$ R$_{\rm r}$ spin-space and real-space rotations. {\bf s}({\bf k}) refers to the microscopic momentum-dependent expectation value of spin. ${\bf d}^{\rm A/B}_{l=1}({\bf k})$	labels the momentum-dependent A/B-phase superfluid order parameter vector in the spin space.
}
\label{fig_He_PI_AM}
\end{figure}

The inherent role of the crystal lattice in the altermagnetic ordering, and the absence of it in superfluid  $^3$He and the Pomeranchuk instability theories, underpins the principal distinction in the respective microscopic ordering mechanisms. In superfluid  $^3$He and in the spin-channel  Pomeranchuk instabilities, the many-body interactions in the Hamiltonian of the fermionic system, together with Pauli exclusion principle, render the momentum-space Fermi surface of the metallic fluid unstable. In the former case, and within a narrow window of low temperatures and high pressures, this results in a spin-triplet Cooper-pairing accompanied by a formation of an anisotropic quasiparticle-excitation gap\cite{Leggett1975} (Fig.~\ref{fig_He_PI_AM}a). In the case of the spin-channel higher-partial-wave Pomeranchuk instabilities, which so far have been experimentally elusive, the correlation-induced instability is predicted to lead to a spin-dependent anisotropic Fermi-surface distortion\cite{Hirsch1990,Oganesyan2001,Wu2004,Wu2007} (Fig.~\ref{fig_He_PI_AM}b).  In contrast, the altermagnetic ordering mechanism is applicable to both metallic and insulating systems. Here the many-body Coulomb interactions and Pauli exclusion principle conspire with the single-particle crystal potential to generate an anisotropic spin density distribution in the crystal, reflected in the momentum space in the corresponding spin-polarized anisotropically distorted energy iso-surfaces (Fig.~\ref{fig_FM_AM_AFM_AAM}b,\ref{fig_He_PI_AM}c)\cite{Smejkal2020,Smejkal2021a}.    

The predicted physical systems potentially hosting the spin-channel higher-partial-wave Pomeranchuk instabilities included heavy fermions, cold atoms, or  $^3$He\cite{Wu2004,Wu2007}. The identified distinct mechanism of the altermagnetic ordering directed the experimental search towards a different class of condensed-matter systems\cite{Smejkal2021a,Smejkal2022a}, which has recently materialized in the spectroscopic and microscopic confirmation of altermagnetism in common binary compounds like MnTe or CrSb\cite{Krempasky2024,Lee2024,Osumi2024,Hajlaoui2024,Chilcote2024,Amin2024,Reimers2024,Yang2024,Ding2024,Zeng2024,Li2024,Lu2024}.

We conclude this opening section by brief comments on dipolar-coupling or spin-orbit-coupling terms originating from the relativistic Dirac equation, which connect the spin-space and real-space reference frames.  As single-particle terms in the Hamiltonian, they are not responsible for the spontaneous ordering. They can remove, however, the degeneracy of different ordered ground states related by symmetries of the non-relativistic Hamiltonian. The resulting magnetic anisotropy energy can facilitate the stability of ordered ground states at discrete orientations of spins which is, e.g., the physical basis of the magnetic memory functionality. Altermagnets can further enrich the landscape of the relativistic spin-dependent effects beyond the phenomenology of conventional magnetic or non-magnetic systems.  Examples are Hall responses which are highly anisotropic and are not generated by internal magnetization or external magnetic field, non-linear altermagnetic relativistic spin-splitting effects in even-parity band structures, or novel topological phases. These phenomena, including their initial experimental exploration, are reviewed, e.g., in  Refs.~\onlinecite{Smejkal2022AHEReview,Smejkal2022a,Bai2024,Jungwirth2024}. 

In the following sections we focus on physics related to the ordering. For this it is desirable to disentangle the spontaneous symmetry breaking in the ordered ground state due to many-body interactions,  from the symmetry breaking in the non-interacting single-particle Hamiltonian arising from the relativistic coupling of the spin-space and the real-space reference frames. To do so, we will consider symmetries of generally different transformations in the spin space and in the real space. This  approach is  analogous to symmetry theories  earlier employed in the studies of superfluid $^3$He and unconventional superconducting phases\cite{Leggett1975,Vollhardt1990,Annett1995,Wolfle1999,Tsuei2000,Leggett2003,Houzet2012,Brison2021,Volovik2022}, as well as of the Pomeranchuk Fermi-liquid instabilities\cite{Akhiezer1978,Hirsch1990,Marchenko1991,Gorkov1992,Oganesyan2001,Kivelson2003,Wu2004,Wu2007,Chubukov2009,Alexandradinata2010,Fischer2011,,Kiselev2017,Wu2018,Klein2019}. A spin-group theory, which systematically applies the approach of uncoupled spin space and real space to the symmetry classification of spin arrangements on crystals\cite{Smejkal2021a,Smejkal2022a,Litvin1974,Litvin1977,Mazin2021,Liu2021,Smejkal2022GMR,Betancourt2021,Reichlova2024,Hariki2023,McClarty2024,Smolyanyuk2024,Shinohara2024,Watanabe2024,Jiang2024,Chen2024,Schiff2023,Xiao2024}, led to the recent delineation of altermagnetism as a distinct third collinear magnetic phase, separate from conventional ferromagnetism and antiferromagnetism.\cite{Smejkal2021a}

The discussion below is organized in the following sections.  In Sec.~\ref{AM} we introduce the symmetry classification and microscopic physics of altermagnetism. Sec.~\ref{spin-independent} places ordered phases preserving the spin-space rotation symmetry in a two-parameter space of interaction strength and conduction type. The discussion serves as a background reference for the follow-up Secs.~\ref{He-AM}  and \ref{PI-AM} on phases which spontaneously break both the spin-space and the real-space rotation symmetries. These two sections reflect on the analogies and distinctions of altermagnetism as compared to superfluid $^3$He and ordered phases generated by the Pomeranchuk Fermi-liquid instabilities, respectively. In Sec.~\ref{beyond} we  extend the comparison beyond the collinear altermagnetic ordering by including non-collinear compensated spin arrangements on crystals (Fig.~\ref{fig_FM_AM_AFM_AAM}d and Fig.~\ref{fig_He_PI_AM}c-f). Finally, we briefly summarize our Perspective in Sec.~\ref{summary}.

\subsection{Altermagnetism}
\label{AM}

The collinear altermagnetic ordering is exclusively and unambigously delineated by the spin-group symmetries\cite{Smejkal2021a}. The non-relativistic many-body interacting Hamiltonian has the spin-group symmetry ${\bf Z}_2^{T}\times{\rm SO(3)}\times {\bf G}$, where ${\bf Z}_2^{T}$ contains the time-reversal ($T$) symmetry,  SO(3) is a group of all continuous spin-space rotations, and  {\bf G} are the crystallographic point groups. 

In the altermagnetically ordered ground state,  the spin-group symmetry is spontaneously lowered to\cite{Smejkal2021a} ${\bf Z}_2^{C_2T}\ltimes{\rm SO(2)}\times ([E\parallel H] +[C_2\parallel {\bf G}-{\bf H}])$. Here the $C_2T$ symmetry in ${\bf Z}_2^{C_2T}$ combines $T$ with a two-fold spin-space rotation $C_2$ around an axis orthogonal to the collinearity axis of spins, SO(2) is a group of continuous spin-space rotations around the collinearity axis,  $E$ is the spin-space identity, {\bf H} is a halving subgroup of  {\bf G}, and ${\bf G}-{\bf H}$ contains proper or improper crystal rotations and does not contain the parity (inversion) transformation. (Note that ${\bf Z}_2^{C_2T}\ltimes{\rm SO(2)}$ is a semidirect product since ${\bf Z}_2^{C_2T}$ is not a normal subgroup of ${\bf Z}_2^{C_2T}\ltimes{\rm SO(2)}$.)

The collinear altermagnetic ground state thus breaks the $T$-symmetry, while it preserves the $C_2T$ symmetry. It also breaks the spin-space SO(3) symmetry, while preserving the SO(2) symmetry. Finally, the altermagnetic ground state  breaks the symmetries of ${\bf G}-{\bf H}$, while it  preserves  the symmetries of {\bf H}  and symmetries combining transformations from ${\bf G}-{\bf H}$ with the spin-space $C_2$ rotation. The latter symmetries protect the zero net magnetization of the compensated collinear altermagnetic ordering.  

In summary, the altermagnetic spin groups explicitly describe the spontaneous breaking of both the spin-space and the real-space rotation symmetries. In contrast, collinear ferromagnetism is delineated by spin groups  ${\bf Z}_2^{C_2T}\ltimes{\rm SO(2)}\times [E\parallel G]$, explicitly showing the preserved rotation symmetries of the parent crystallographic point-group\cite{Smejkal2021a}. 

In the momentum space of the altermagnetic phase, the energy iso-surface of a given spin is distorted, breaking the symmetries of ${\bf G}-{\bf H}$  and preserving the {symmetries of \bf H}. The  iso-surfaces corresponding to opposite spins are mutually related by the symmetries of ${\bf G}-{\bf H}$. Near the $\boldsymbol\Gamma$-point, the electronic structure is spin-degenerate at 2, 4, or 6 nodal surfaces in the 3D Brillouin zone, depending on the spin-symmetry group. On either side of the nodal surface, the sign of the spin-polarization alternates. Correspondingly, the altermagnetic ordering can be of $d$, $g$, or $i$-wave type\cite{Smejkal2021a}. Note that, in the momentum space, the electronic structure has the parity symmetry, regardless of whether in the position space of the crystal lattice the parity symmetry is present or broken. This is due to the $C_2T$ symmetry which (in combination with SO(2)) acts as parity symmetry in the momentum space\cite{Smejkal2021a}. We also remark that the crystallographically constrained possible discrete real-space rotation transformations  restrict the allowed even-parity-wave altermagnetism to $d$, $g$ or $i$-wave. 

Numerous material candidates of not only $d$-wave, but also $g$-wave and $i$-wave altermagnets have been identified using the spin-group classification, supported by density-functional-theory calculations\cite{Smejkal2021a,Smejkal2022a,Guo2023b,Xiao2023}.  The initial experimental spectroscopic confirmations were reported  in $g$-wave altermagnets MnTe and CrSb\cite{Krempasky2024,Lee2024,Osumi2024,Hajlaoui2024,Chilcote2024,Amin2024,Reimers2024,Yang2024,Ding2024,Zeng2024,Li2024,Lu2024}. A spectroscopic evidence of $d$-wave altermagnetism has been recently reported in RbV$_2$Te$_2$O\cite{Zhang2024c} and KV$_2$Se$_2$O\cite{Jiang2024a}.
%We will not continue in this Perspective with the systematic symmetry-group description of spin-arrangements on crystals, only point out here that there is more spin point groups of the collinear altermagnetic order (37) than ferromagnetic order (32)\cite{Smejkal2021a}. 

%The collinearity also implies that, in the momentum space, the electronic structure of altermagnets has the parity symmetry, regardless of whether in the position space of the crystal lattice the parity symmetry is present or broken\cite{Smejkal2021a}. This is because all collinear magnets have an additional symmetry combining a two-fold spin-space ration around an axis orthogonal to spins combined with time reversal. In the momentum space of the collinear magnets, this combined transformation acts as parity transformation. 

\begin{figure}[h!]
	\centering
	\includegraphics[width=1\linewidth]{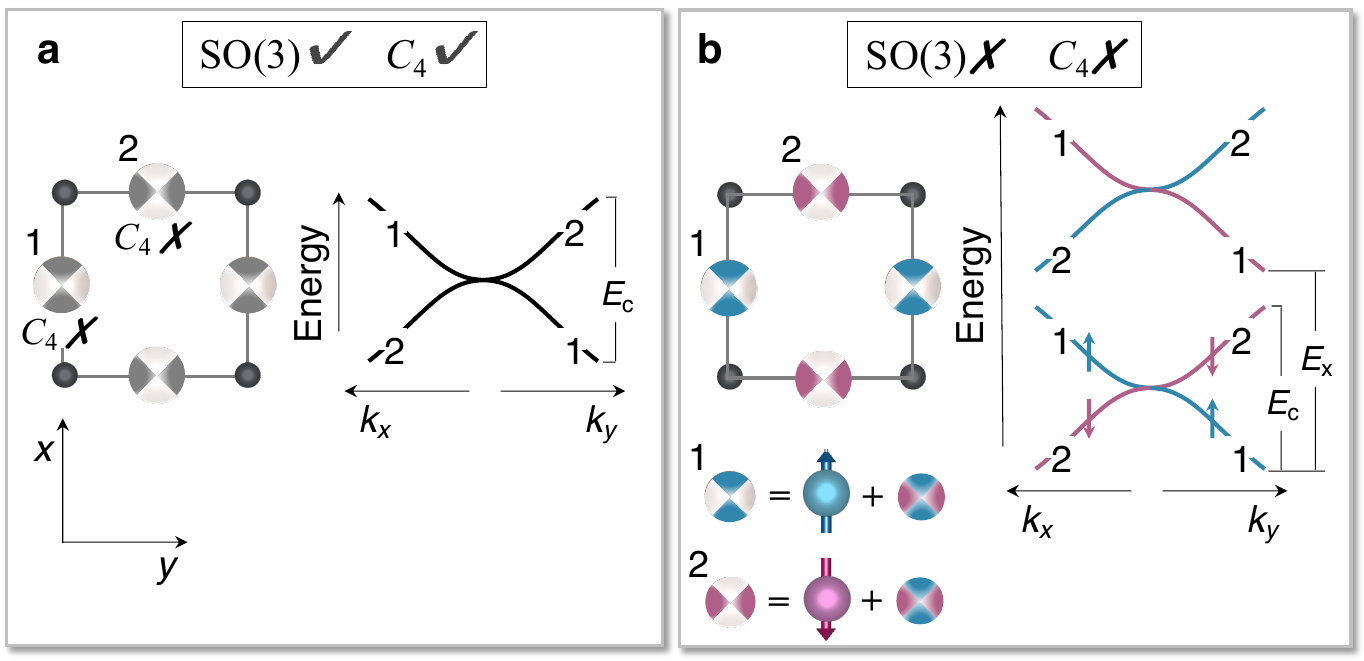}
	\caption{
	\textbf{Model dissection of  altermagnetic symmetry breaking and microscopic ordering mechanism.}
(a) Left: A model square Lieb lattice in the normal phase with the SO(3) spin-space symmetry. From the real-space symmetry of the crystallographic point group we specifically highlight the global $C_4$ rotation symmetry, which is locally broken on sites 1 and 2. The anisotropic charge distribution of orbitals on sites 1 and 2 is highlighted by the anisotropic shading on the sites. Right: Cartoon of a model band structure in the normal phase assuming  two pairs of spin-degenerate bands, dominated by orbitals form sites 1 and 2, respectively. Individually, the bands reflect the locally broken $C_4$ symmetry, and the scale of their mutual orbital-splitting ($E_c$) is determined by the single-particle crystal potential. (b) Same as (a) for the ordered altermagnetic phase spontaneously breaking the SO(3) spin-space symmetry and the global real-space $C_4$ symmetry. Blue and red colors (arrows) depict opposite spin polarization. In the band structure, the spin degeneracy is lifted by the many-body exchange interaction ($E_x$), while the magnitude and momentum-dependence of the spin-splitting of nearby bands copies the orbital band-splitting in the normal phase. Bottom-left of the panel shows a decomposition of the local anisotropic spin density on sites 1 and 2 into a dipole (marked by arrow) and a higher-partial-wave ($d$-wave) spin-density component.}
\label{fig_AM}
\end{figure}

In the remaining part of this section we leave the formal spin-group theory and, following Refs.~\onlinecite{Smejkal2021a, Smejkal2022a}, we dissect the altermagnetic symmetry breaking and microscopic ordering mechanism using a model altermagnet.  Here we choose for the model a square Lieb lattice\cite{Mazin2023a,Antonenko2024}, whose cartoons of the crystal structure in the position space  and of the energy spectrum in the momentum-space  are shown in Fig.~\ref{fig_AM}. In the normal phase, the model crystal has the continuous SO(3) spin-space symmetry and the crystallographic point-group symmetry, from which we explicitly highlight the four-fold rotation $C_4$. These are symmetries of the whole crystal (unit cell) and, correspondingly, the electronic band structure in the normal phase is spin-degenerate and four-fold symmetric in the momentum space. However, still in the normal phase, the $C_4$ symmetry is already locally broken on sites 1 and 2 (Fig.~\ref{fig_AM}a). The model energy spectrum shows  two pairs of spin-degenerate bands, where one pair is dominated by orbitals from site 1  and the other pair by orbitals from site 2 (Fig.~\ref{fig_AM}a). The bands are anisotropic, reflecting the locally broken $C_4$ site-symmetry, and the energy scale of their orbital splitting ($E_c$ in Fig.~\ref{fig_AM}a) is determined by the single-particle crystal potential.  

The altermagnetic order spontaneously breaks the SO(3) spin-space symmetry. In addition, the  $C_4$ symmetry, which was already broken in the normal phase locally on the two sites, becomes a spontaneously globally broken symmetry of the crystal unit cell in the altermagnetic phase. The ordered phase preserves the symmetry combining the real-space $C_4$ rotation with the spin-space $C_2$ rotation. These symmetry features are highlighted by the schematic illustration of the anisotropic spin density and its decomposition into the dipole and the higher-partial-wave ($d$-wave) components (Fig.~\ref{fig_AM}b). 

Remarkably, the higher-partial-wave component of the spin density is ordered ferroically on sites 1 and 2. In analogy to ferroically ordered dipoles in conventional ferromagnetism (Fig.~\ref{fig_FM_AM_AFM_AAM}a),  this leads to broken time-reversal symmetry of the momentum-space electronic structure, and to corresponding time-reversal symmetry breaking electronic responses\cite{Smejkal2020}. The distinction from ferromagnetism is the higher partial-wave nature of the ferroic component  in altermagnetism, which corresponds to the higher partial-wave symmetry of anisotropic exchange interactions in the crystal lattice\cite{Smejkal2023}.

The spin polarization in the momentum-space electronic structure  also reflects the higher-partial-wave ($d$-wave) symmetry of the ferroic component of the anisotropic spin density in the position space of the crystal (Fig.~\ref{fig_AM}b). The spin-degeneracy of energy bands in the momentum-space is lifted in the altermagetic phase by the many-body interaction ($E_x$ exchange-energy scale in Fig.~\ref{fig_AM}b). Remarkably, however, the  spin splitting of nearby bands in the model band structure copies the anisotropic momentum-dependence  and the magnitude of the orbital band splitting in the normal phase ($E_c$ crystal-potential scale in Figs.~\ref{fig_AM}a,b)\cite{Smejkal2021a, Smejkal2022a}. This further illustrates that the spontaneous breaking of the global $C_4$ symmetry in the ordered phase is prearranged by the locally broken $C_4$ site symmetry in the normal-phase.

The above link from the orbital splitting in Fig.~\ref{fig_AM}a to the spin splitting in Fig.~\ref{fig_AM}b is a cartoon example illustrating the microscopic mechanism of the altermagnetic ordering in which the single-particle crystal potential conspires with  the many-body electron-electron (exchange) interaction to form the unconventional phase spontaneously breaking both the spin-space and the real-space rotation symmetries. The mechanism is robust thanks to the typical $\sim$eV scales of both the crystal-potential and the electron-electron interactions, as confirmed in several altermagnetic candidate materials by density-functional theory calculations\cite{Smejkal2021a, Smejkal2022a}. (Note that, for comparison, the relativistic dipolar or spin-orbit interactions are typically on a $\sim$meV scale, unless involving heavy-element orbitals or extreme magnetic fields.) 

Besides the robustness, altermagnetism is also predicted to be abundant. Its microscopic ordering mechanism  applies  to both metallic and insulating systems\cite{Smejkal2021a, Smejkal2022a}. In comparison, ferromagnetism is mostly realized in metals. Moreover, from the symmetry perspective, out of the 122/1421 spin point/space groups of all collinear spin arrangements on 3D crystals, 37/422 correspond to the altermagnetic order, in comparison to 32/230 corresponding to the conventional ferromagnetic order (and 53/769 to  conventional  antiferromagnetic order)\cite{Smejkal2021a}.

\subsection{Ordered phases preserving spin-space rotation symmetry}
\label{spin-independent}
\begin{figure}[h!] 
	\centering
	\includegraphics[width=1\linewidth]{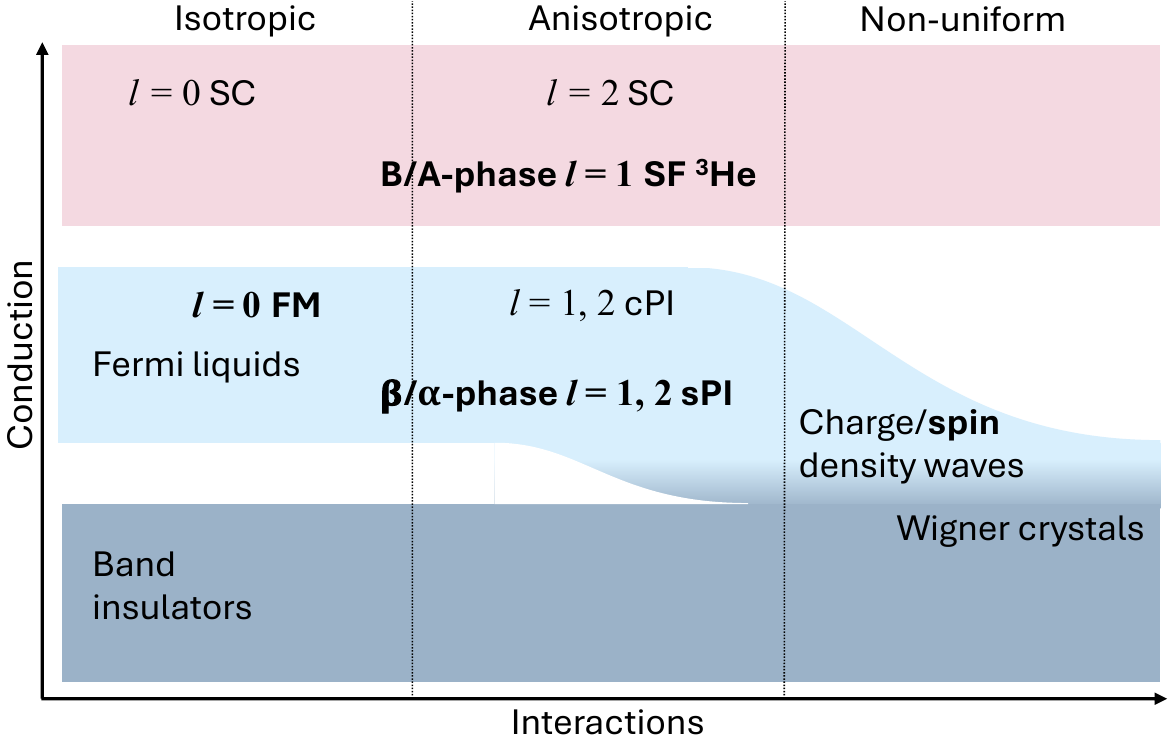}
	\caption{
	\textbf{Illustrative diagram of the landscape of other condensed-matter phases.}
Horizontal axis corresponds to increasing strength of interactions. Insulating, metallic and superconducting/superfluid phases are separated along the vertical axis. Bold symbols highlight phases with spontaneously broken spin-space rotation symmetry. FM refers to ferromagnets, SC to superconductors, SF to superfluid, and cPI and sPI to charge and spin-channel Pomeranchuk instabilities, respectively.
}
\label{fig_int_cond}
\end{figure}

We now turn  to the comparison of altermagnetism with other condensed-matter phases, summarized in Fig.~\ref{fig_int_cond}. In this section we briefly recap, as a reference, phases which preserve the spin-space rotation symmetry. Painted with a broad brush, we can make the following (non-exhaustive) classification of phases without spin ordering by the conduction type and interaction strength, as shown in Fig.~\ref{fig_int_cond}. Starting from the bottom-left of the diagram, we have band insulators whose physics is described by an effective single-particle  band picture featuring an energy gap separating completely filled bands from empty bands. The band-insulating phases are produced by quantum-interference effects of electrons in the periodic potential of the  crystal lattice. 

Metallic phases feature a Fermi surface separating the occupied and empty electronic states. Landau's Fermi liquid theory\cite{Landau1957,Vignale2022} provides an elegant explanation why, in many metals,  excited states near the Fermi surface can be represented by weakly interacting fermion quasiparticles whose  lifetime becomes infinite when approaching the Fermi surface. This is because the scattering phase-space for the excited states near the Fermi surface is drastically limited by the Pauli exclusion principle. Normal metals falling into this weak-interaction Fermi-liquid regime are depicted in the middle-left part of the diagram in Fig.~\ref{fig_int_cond}. 

\begin{figure}[h!]
	\centering
	\includegraphics[width=1\linewidth]{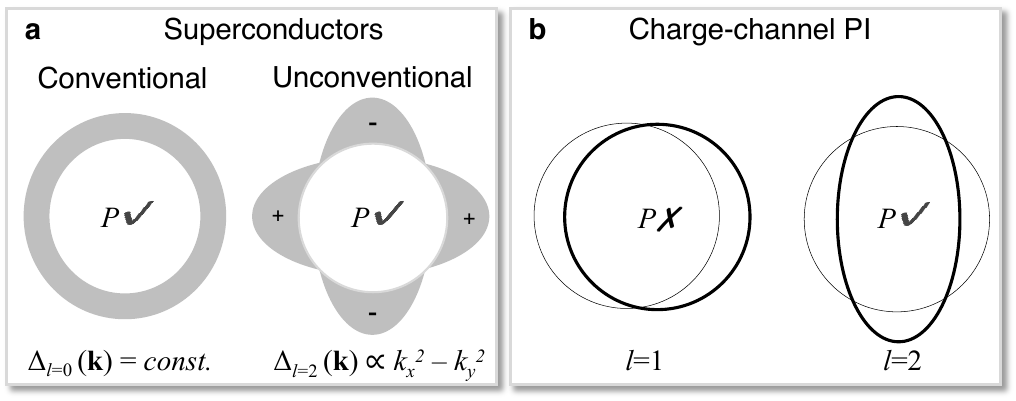}
	\caption{
	\textbf{Ordered phases preserving spin-space rotation symmetry.}
Left panel: Even-parity order parameters of conventional $s$-wave ($l=0$, left) and unconventional $d$-wave ($l=2$, right) superconductors. Right panel: Odd parity $p$-wave ($l=1$, left) and even-parity $d$-wave ($l=2$, right) charge-channel Pomeranchuk instabilities of the Fermi surface. Thin lines show normal-phase Fermi surfaces of an isotropic Fermi liquid.}
\label{fig_SC_PI}
\end{figure}

At sufficiently low temperatures, an arbitrarily weak attractive interaction (e.g. mediated by phonons) between the quasiparticles near the Fermi surface leads to the formation of Cooper pairs of opposite-spin and opposite-momentum quasiparticles, corresponding to a spin-singlet ($S=0$) $s$-wave  ($l=0$) pairing (Fig.~\ref{fig_SC_PI}a). The occupied and empty fermionic states get separated by an excitation gap when the Cooper pairs condense into the conventional BCS superconducting state\cite{Bardeen1957}, which we correspondingly placed in the top-left part of Fig.~\ref{fig_int_cond}. 

The valence-band energy iso-surfaces in the band-insulators, the Fermi surface in the normal metals, or the order parameter in the $s$-wave superconductors (Fig.~\ref{fig_SC_PI}a) preserve the spin-space and the real-space rotation symmetries. Note that the superconducting/superfluid ordering spontaneously breaks the U(1) gauge (particle-conservation) symmetry.

We now move to the unconventional superconductivity\cite{Wolfle1999,Tsuei2000} in the strongly-interacting part of the diagram in Fig.~\ref{fig_int_cond}. Because of strong short-range repulsive interactions, electrons in the Cooper pairs favor anisotropic pairing with relative angular momentum $l>0$. This breaks the real-space rotation symmetry. Unconventional $d$-wave ($l=2$) cuprate superconductors  belong to this  class (Fig.~\ref{fig_SC_PI}a). The symmetric orbital part  of the pairing function dictates, by the Pauli exclusion principle, an antisymmetric spin-singlet ($S=0$) part of the pairing function. The $d$-wave superconducting phase, like the conventional $s$-wave superconductivity, thus  preserves the spin-space rotation symmetry.

Wigner crystal is another ordered phase in the stongly-interacting part of the diagram in Fig.~\ref{fig_int_cond}. It is stabilized when  the interaction energy dominates the kinetic energy at sufficiently low electronic densities. Ordering in this insulating phase is characterized by a non-uniformity, i.e., by a spontaneous breaking of the translation symmetry. Charge-density waves, stripe states or electronic smectic liquid crystals are other examples  of correlated non-uniform ordered phases\cite{Fradkin2007}. They can be insulating or metallic.

Finally, we arrive at the central part of the diagram in Fig.~\ref{fig_int_cond} corresponding to an intricate metallic intermediate-interaction regime. Coming from the right, the corresponding ordered phases can be viewed as melted charge-density waves\cite{Kivelson1998,Fradkin1999,Rezayi1999,Jungwirth1999a,Fradkin2007}, recovering the translation symmetry, but not the real-space rotation symmetry\cite{Fradkin2007}. Coming from the left, they can be described as Pomeranchuk Fermi-liquid instabilities in the $l>0$ charge channel, characterized by distortions of the Fermi surface and the corresponding spontaneous breaking of the real-space rotation symmetry (Fig.~\ref{fig_SC_PI}b)\cite{Halboth2000,Oganesyan2001,Fradkin2007}. Simultaneously, they preserve the uniformity (translation symmetry) of the Fermi liquid and the spin-space rotation symmetry.
%Note that in this article we use the term nematic electronic liquid-crystal phases in a generalized sense to underline the spontaneous breaking of the real-space rotation symmetry not only in the even-parity $l=2$ case, which has been studied recently in various quantum materials, but in all even and odd-parity $l>0$ phases\cite{Wu2004,Wu2007}. 

We will elaborate on the  framework of the Pomeranchuk Fermi-liquid instabilities in more depth in the next section where we move to the discussion of the spin-ordered phases\cite{Hirsch1990,Oganesyan2001,Kivelson2003,Wu2004,Wu2007,Fischer2011,Kiselev2017,Wu2018,Klein2019}.  Here we conclude the overview of the   phases preserving the spin-space rotation symmetry by pointing out that the charge-channel $l=1$ Pomeranchuk instability has been a matter of an on-going discussion throughout the past hundred years. In Fig.~\ref{fig_SC_PI}b, the instability is depicted as a Fermi-surface distortion in the form of a parity-breaking shift. From the 1920's till the 1940's, considered among others by Bloch, Landau or Born\cite{Born1948,Bohm1949,Hoddeson1987,Schmalian2010}, such a Fermi-surface distortion, suggesting a presence of a spontaneous equilibrium current, recurrently appeared as an attempt to explain conventional superconductivity. The theory failed as it violated the first theorem on superconductivity, formulated in the meantime by Bloch himself, showing that the minimum energy state bears no current\cite{Hoddeson1987}. With the other unsuccessful theories at the time,  it led Bloch to his "second theorem" stating that every theory of superconductivity could be disproved\cite{Hoddeson1987}. While the conventional superconductivity was eventually explained by the Cooper-pairing mechanism\cite{Bardeen1957}, whether or not the charge-channel $l=1$ Pomeranchuk instability is physically possible has remained a matter of theoretical research till today\cite{Kiselev2017,Wu2018,Klein2019}. Experimentally, it has remained elusive.

The charge-channel $l=2$ Pomeranchuk instability, leading to an ordered phase referred to as nematic\cite{Halboth2000,Oganesyan2001,Fradkin2007}, has a form of an even-parity anisotropic Fermi-surface distortion (Fig.~\ref{fig_SC_PI}b). The studied physical realizations of this electronic nematic phase include semiconducting 2D electron systems at high magnetic fields in the vicinity of correlated fractional quantum-Hall states, correlated ruthenates  at high magnetic fields in the vicinity of a metamagnetic transition, as well as unconventional superconductors such as cuprates and iron pnictides\cite{Lilly1999,Oganesyan2001,Borzi2007,Fradkin2007,Fradkin2010,Fernandes2014,Lee2018b,Quintanilla2023}. 

Finally, we note that no general limitations have been identified to exclude  $l>2$ Pomeranchuk instabilities. For example,  a five-fold $l=5$ Fermi-surface distortion has been theoretically considered  in a uniform Fermi liquid unconstrained by crystal-lattice symmetries\cite{Quintanilla2008}.  
%**************************************************************************
\subsection{Superfluid $^3$He and altermagnetism}
\label{He-AM}
We now turn the focus to the ordered phases that spontaneously break both the spin-space and the real-space rotation symmetries, and compare them to altermagnetism. Starting from the top of the diagram in Fig.~\ref{fig_int_cond}, the well-established representative\cite{Leggett2003} is superfluid $^3$He. Already since late 1950's, more than a decade before the experimental discovery, theorists were considering extensions of the BCS theory  to the case of charge-neutral spin-1/2 $^3$He atoms. Here the longer-range attractive van der Waals interaction and the  strong short-range repulsion, complemented by the additional effective interaction due to spin fluctuations, favor spin-triplet $p$-wave pairing ($S=l=1$)\cite{Leggett2003}. 

Despite the $p$-wave pairing, the expectation was that $^3$He would remain essentially isotropic in the superfluid phase. Specifically,  the  lowest energy superfluid state originally predicted by the microscopic theory, the so-called B-phase,  is formed by a condensation of Cooper pairs given by a linear combination containing all three $l_z$ values, $|l_z=-1, S_{z'}=1\rangle+|l_z=1, S_{z'}=-1\rangle+|l_z=0, S_{z'}=0\rangle$. (We recall that dipolar and spin-orbit coupling is omitted which we highlight in the pairing  by introducing distinct, unprimed and primed coordinate systems for the real space and the spin space.) Using $l=1$, $l_z=\pm1,0$ spherical harmonics, the B-phase order parameter along the Fermi surface is commonly written as a vector in the spin-space whose components depend on  the momentum unit vector as\cite{Leggett1975,Vollhardt1990,Wolfle1999,Leggett2003,Brison2021,Volovik2022} ${\bf d}^{\rm B}_{l=1}({\bf k})\propto{\bf k}$.  The direction of ${\bf d}^{\rm B}_{l=1}({\bf k})$ varies with {\bf k} but the constant amplitude implies that the quasiparticle excitation gap in the B-phase is open along the whole Fermi surface and has a constant value, i.e., is isotropic (Fig.~\ref{fig_He_PI_AM}d). 

We emphasize that the order-parameter vectors ${\bf d}_{l=1}({\bf k})$ describing superfluid phases of $^3$He, albeit not directly proportional to spin, are constructed to follow the same rotation as the spin quantization axis under spin-space rotations\cite{Leggett1975,Vollhardt1990,Wolfle1999,Leggett2003,Brison2021,Volovik2022}. The order parameter ${\bf d}^{\rm B}_{l=1}({\bf k})$ thus explicitly shows that the B-phase of superfluid $^3$He breaks, individually, the spin-space rotation symmetry SO(3)$_s$ and the real-space rotation symmetry SO(3)$_r$. Simultaneously, it retains an SO(3) symmetry of the same combined rotations in the spin space and the real space, underlying the essentially isotropic nature of the B-phase (Fig.~\ref{fig_He_PI_AM}d). In addition, ${\bf d}^{\rm B}_{l=1}({\bf k})$ shows that the B-phase retains the time-reversal symmetry and breaks the parity symmetry (Fig.~\ref{fig_He_PI_AM}d). 

Consistent with the original theoretical expectation, the experimental phase diagram of superfluid $^3$He is indeed dominated by the B-phase. A major surprise thus was the experimental observation of the A-phase of superfluid $^3$He, occurring within a narrow window of low temperatures and high pressures\cite{Osheroff1972a,Leggett2003}. Its  spin-triplet $p$-wave Cooper pairing function is of the form $|l_z=1, S_{z'}=1\rangle+|l_z=1, S_{z'}=-1\rangle$, containing only one of the three $l_z$ values.  The corresponding order parameter is given by,  ${\bf d}^{\rm A}_{l=1}({\bf k})\propto (0, ik_x - k_y, 0)$.  For all momenta, the vector has only one non-zero component, i.e., is collinear along the Fermi surface. Its amplitude, however, varies with momentum, rendering the A-phase anisotropic with nodes in the quasiparticle excitation gap at $k_x=k_y=0$ (Fig.~\ref{fig_He_PI_AM}a).   There is no remaining SO(3) symmetry in the A-phase which underlines its anisotropic nature. The order parameter ${\bf d}^{\rm A}_{l=1}({\bf k})$, containing only one $l_z$ component, also implies broken time-reversal symmetry in the A-phase (Fig.~\ref{fig_He_PI_AM}a). 

The nodes in the gap function make the A-phase of superfluid $^3$He less favorable than the B-phase with the isotropic nodeless gap function. The surprising and limited occurrence of the anisotropic A-phase of superfluid $^3$He is then explained by a subtle feedback effect of the pairing state on the spin fluctuations which contribute to the attraction forming the Cooper pairs\cite{Anderson1973}. The B-phase has a reduced spin-susceptibility compared to the normal phase and the A-phase. Since the pairing mechanism involves spin fluctuations, this can disfavor the B-phase. The lesson learned from superfluid  $^3$He was that it may require an intricate and subtle interplay of microscopic interactions to realize an ordered phase which spontaneously breaks both the spin-space and real-space rotation symmetries, and where these lead to an anisotropic nodal character of the ordering.

Altermagnets represent a new physical realization of  the ordered phase belonging to the  symmetry class with spontaneously broken spin-space and real-space rotation symmetries. It also shares with the A-phase of superfluid $^3$He the anisotropic nodal character (Fig.~\ref{fig_He_PI_AM}a,c). In contrast to $^3$He, however, the altermagnetic ordering was theoretically anticipated prior to the experimental discovery\cite{Smejkal2022a}. Moreover,  the expectation was based not only on a theory applied to a specific physical system, but on the general spin-group classification of spin arrangements on crystals\cite{Smejkal2021a}. As a result, numerous altermagnetic candidates have been identified, many of which order at ambient conditions, and not only in 3D inorganic materials\cite{Smejkal2021a, Smejkal2022a,Guo2023b,Xiao2023}, but also in 2D \cite{Smejkal2022a,Smejkal2022GMR,Ma2021,Egorov2021,Brekke2023,Cui2023,Chen2023b,Mazin2023a,Sodequist2024} and organic crystals\cite{Naka2019,Ferrari2024}. The initial experimental demonstrations by momentum-space spectroscopic measurements have been performed in room-temperature altermagnetic materials MnTe and CrSb\cite{Krempasky2024,Lee2024,Osumi2024,Hajlaoui2024,Chilcote2024,Reimers2024,Yang2024,Ding2024,Zeng2024,Li2024,Lu2024}, representing simple binary compounds readily available in stable high-quality bulk or thin-film forms. The spectroscopy has been complemented in MnTe by position-space vector-imaging and control of the altermagnetic ordering from micron-scale single-domain states to nano-scale domain walls and topological vortices\cite{Amin2024}. 

The robustness of altermagnetism stems from the specific microscopic physics of ordering. For superfluid $^3$He, we mentioned above the key role of subtle effects of the effective attractive interaction via spin fluctuations. The effective interactions and, in general, the vicinity of other (fluctuating)  phases of the interacting Fermi fluid\cite{Schofield2009}, has been a common theme considered across the field of the anisotropic phases, including the unconventional superconductors and the charge-channel Fermi-liquid instabilities, discussed in the previous section. Altermagnets stand apart here. Although intriguing physics may arise from an interplay of the altermagnetic phase with other order parameters of the interacting electrons\cite{Smejkal2022a,Jungwirth2024}, altermagnetic ordering is not based on this interplay. As described in Sec.~\ref{AM}, altermagnetism is primarily stabilized, apart from the internal many-body exchange interaction, by the external static single-particle potential of the underlying crystal lattice\cite{Smejkal2020,Smejkal2021a,Smejkal2022a,Jungwirth2024}. The crystal-potential energy scale tends to be strong which adds to the robustness of altermagnets in metallic and insulating materials over a broad range of electron-electron interaction strengths.

%P in AMs vs. broken P in 3He. Let's if it's good to bring it up here or later.....

%**************************************************************************
\subsection{Spin-channel Pomeranchuk instabilities and altermagnetism}
\label{PI-AM}

In this section we proceed by comparing altermagnetism to theoretically studied spin-ordered phases generated by spin-channel Pomeranchuk instabilities of the momentum-space Fermi surface. They are placed in the  part of the diagram in Fig.~\ref{fig_int_cond} corresponding to metallic conduction and  intermediate interaction strengths. Together with the charge-channel Pomeranchuk instabilities, mentioned in Sec.~\ref{spin-independent}, their phenomenological description can be put under the common umbrella based on Landau Fermi-liquid theory. The theory considers two-body interactions between the quasiparticles.  In a free space and for quasiparticles near the Fermi surface, the two-body interaction depends only on the angle~$\theta$ between  the linear momenta of the quasiparticles. Accordingly, it can be expanded in series of ($l$, $l_z=0$) spherical harmonics, correspondingly called angular-momentum $l$-channels\cite{Landau1957,Vignale2022}.  The $l=0$ component corresponds to an isotropic interaction while the $l>0$ components describe anisotropic  interactions. The prefactors of the expansion are the phenomenological Landau  parameters, $F^{\rm c(s)}_l$. Here the superscript c(s) labels the charge (spin) channel interaction given by the sum (difference) of interactions of the same-spin and opposite-spin quasiparticles. 

Pomeranchuk derived\cite{Pomeranchuk1959} a general form of the static susceptibility in each charge, spin and $l$-channel, finding it to be proportional to $1/(1+F^{\rm c(s)}_l)$.  (Here $F^{\rm c(s)}_l$ are conveniently normalized\cite{Vignale2022}.) The susceptibility diverges at $F^{\rm c(s)}_l=-1$, signalling the Pomeranchuk instability of the Fermi liquid\cite{Pomeranchuk1959}. The spin-channel $l=0$ ($s$-wave) instability corresponds to the conventional  metallic ferromagnetic phase\cite{Landau1957,Pomeranchuk1959}. 
%The spin-degenerate Fermi surface of the normal state is split in the ordered state into  larger and smaller Fermi surfaces, one corresponding to spin-up electrons and the other one to spin down electrons. This breaks the spin-space rotation symmetry. For the Fermi liquid in a free space, the split Fermi surfaces have an isotropic shape, respecting the continuous real-space rotation symmetry. On a lattice, the ferromagnetic ground state preserves the discrete crystallographic point-group symmetries (Fig.~\ref{fig_FM_AM_AFM_AAM}). 
The anisotropically distorted spin-degenerate Fermi surfaces, mentioned in Sec.~\ref{spin-independent} and illustrated in Fig.~\ref{fig_SC_PI}b, correspond to the charge-channel $l=1$ ($p$-wave) \cite{Born1948,Bohm1949,Landau1957,Pomeranchuk1959,Hoddeson1987} and  $l=2$ ($d$-wave)\cite{Halboth2000,Oganesyan2001,Fradkin2007} Pomeranchuk instabilities, respectively.  Their spin-channel counterparts are illustrated in Figs.~\ref{fig_He_PI_AM}b,e. 

Starting from  $l=1$, referred to as spin nematic in the early literature\cite{Marchenko1991,Gorkov1992}, and later identified as a generalized electronic liquid crystal\cite{Wu2007}, there are two types labeled in Refs.~\onlinecite{Wu2004,Wu2007} as the $\alpha$-phase  and the $\beta$-phase (Fig.~\ref{fig_He_PI_AM}b,e), in analogy to the A-phase and the B-phase of superfluid $^3$He (Fig.~\ref{fig_He_PI_AM}a,d). 

In the $\beta$-phase, the Fermi surface spin-splits into larger and smaller surfaces, where the shape of each surface is otherwise undistorted\cite{Akhiezer1978,Marchenko1991,Wu2004,Wu2007}. This is reminiscent of the $l=0$ ($s$-wave) ferromagnetic instability, but there  are key symmetry differences. First, the $\beta$-phase  $l=1$ Pomeranchuk instability shares with the B-phase of superfluid $^3$He the time-reversal symmetry and the broken parity symmetry. In contrast, the ferromagnetic instability breaks time-reversal symmetry and preserves the parity symmetry. Second, in analogy to the spin-space order parameter vector ${\bf d}^{\rm B}({\bf k})\propto{\bf k}$  in the B-phase of superfluid $^3$He, the $l=1$ $\beta$-phase has a momentum dependent spin texture on the energy iso-surface, where the direction of spin  depends on the direction of momentum such that the spin winds once along the Fermi surface. As a result, the $l=1$ $\beta$-phase breaks, individually, both the spin-space and real-space rotation symmetries, but preserves a symmetry combining the same rotation transformations in the spin space and the real space, rendering the phase essentially isotropic. This is analogous to the B-phase of superfluid $^3$He. In contrast, the $l=0$ ferromagnetic Pomeranchuk instability is isotropic because it only breaks the spin-space rotation symmetry, but preserves the real-space rotation symmetry.

The $\alpha$-phase generated by the $l=1$ spin-channel Pomeranchuk instability is illustrated in Fig.~\ref{fig_He_PI_AM}b as spin-split Fermi surfaces shifted along one direction for one spin and the opposite direction for the opposite spin\cite{Hirsch1990,Marchenko1991,Wu2004,Wu2007}. The phase is more reminiscent of the A-phase of superfluid $^3$He in the following sense. Besides breaking the spin-space and real-space rotation symmetries, and in analogy to ${\bf d}^{\rm A}_{l=1}({\bf k})\propto (0, ik_x - k_y, 0)$ in the A-phase of superfluid $^3$He, spins in the $\alpha$-phase are collinear on the Fermi surfaces, and the phase is anisotropic with spin-degenerate nodes. However, the $l=1$ $\alpha$-phase Pomeranchuk instability, like the $l=1$ $\beta$-phase but unlike the A-phase of superfluid $^3$He, retains the time-reversal symmetry. 

In the spin-channel Pomeranchuk instabilities, the time-reversal symmetry is broken for spin-ordered phases generated by the $l=2$ (even $l$) instability, i.e., in phases retaining the parity symmetry\cite{Oganesyan2001,Wu2007}. The $\alpha$-phase  $l=1$ and $l=2$ Pomeranchuk instabilities  both share with  the A-phase of superfluid $^3$He the broken spin-space and real-space rotation symmetries and the anisotropic collinear and nodal character of ordering (Fig.~\ref{fig_He_PI_AM}a,b). However, they depart from the A-phase of superfluid $^3$He in one of the two discrete symmetries (Fig.~\ref{fig_He_PI_AM}a,b): The A-phase of superfluid $^3$He spontaneously breaks both the time-reversal and parity symmetry, while the $l=1$  $\alpha$-phase Pomeranchuk instability retains the time-reversal symmetry and breaks the parity symmetry. {\em Vice versa}, the $l=2$  $\alpha$-phase Pomeranchuk instability breaks the time-reversal symmetry and retains the parity symmetry.  (Recall here that the even-parity superfluid/superconducting phases with even-$l$ Cooper-pairing are spin-singlet and do not break the spin-space rotation symmetry).

A cartoon representation of momentum-space spin-dependent energy iso-surfaces of a $d$-wave altermagnet (Fig.~\ref{fig_He_PI_AM}c) illustrates the symmetry analogy to the $l=2$   $\alpha$-phase Pomeranchuk instability (Fig.~\ref{fig_He_PI_AM}b).  However, the microscopic mechanism that stabilizes the altermagnetic order in a broad range of materials, including also insulators, is principally distinct from the Pomeranchuk instabilities. In the latter case, the interactions have to be strong enough for the corresponding Landau parameter $F^{\rm s}_l$ to reach the critical value of the Pomeranchuk instability in the given $l$-channel while, simultaneously, the interactions have to be fine-tuned to avoid the instability in another $l$-channel. In particular, eliminating a pre-emptive $l=0$ ferromagnetic instability was recognized as one of the key challenges for realizing the spin-ordered phases by $l>0$ Pomeranchuk instabilities\cite{Hirsch1990,Wu2004,Wu2007,Fischer2011}. Apart from the isotropic ferromagnetic ordering, a spin-density-wave  phase can compete with the $l>0$ Pomeranchuk instabilities from the non-uniform side\cite{Hirsch1990} of the diagram in Fig.~\ref{fig_int_cond}. To date, the spin-channel $l>0$ Pomeranchuk instabilities  (without an interplay of strong spin-orbit coupling) have remained largely as theoretical concepts.  

We also point out that in theories of the spin-channel Pomeranchuk instabilities, the crystal-lattice tends to enter only indirectly via a modification of the one-body momentum-dependent energy dispersion of the quasiparticle states. For example, an anisotropic quasiparticle-dispersion and corresponding Fermi-surface nesting effects in suitable lattice models were considered as a possible mechanism for favoring the  $l=1$ $\alpha$-phase instability against the $l=0$ ferromagnetic instability\cite{Hirsch1990}. Alternatively, effects beyond quadratic quasiparticle-dispersion were shown to determine whether, for a critical value of a given  Landau parameter $F^{\rm s}_{l>0}$, the corresponding ordering will tend to be of the $\alpha$-phase or the $\beta$-phase\cite{Wu2007}. 
%We also note that the absence of lattice-symmetry constraints enables, e.g., an $l=5$ instability. 

In contrast, as discussed in Sec.~\ref{AM}, the crystal lattice plays an inherent role in the altermagnetic ordering. As a result, the altermagnetic spin-group symmetry  protects zero net magnetization, i.e., excludes a  ferromagnetic component.  Moreover, the collinearity of the spin arrangement in altermagnets in both the position space of the crystal lattice and the reciprocal momentum space excludes the spin-textured type of ordering, which in the Pomeranchuk instabilities corresponds to the $\beta$-phase. 
%Lattice-symmetry constraints restrict altermagnetism to d, g, or i-wave ordering.

To further highlight the distinction between microscopic ordering mechanisms of altermagnetism and the Pomeranchuk instabilities, we compare minimal models associated with each scenario. We focus for concreteness on  $d_{x^2-y^2}$-wave symmetry. In the Pomeranchuk scenario, the electronic dispersion in the spin-ordered state near the $\boldsymbol\Gamma$ point becomes $E_\sigma(\boldsymbol{k}) = E^0(\boldsymbol{k}) + \phi (k_x^2 - k_y^2)\sigma$, where $\sigma=\pm 1$ is the spin index, $\phi\neq 0$ in the ordered state, and $E^0(\boldsymbol{k}) \sim k^2$. For this ordering-transition to take place, the Landau parameter $F^{\rm s}_{2}$ must overcome a threshold value while all $F^{\rm{s}}_{l\neq2}$ must remain below their critical values. 

This scenario, however, is very challenging to be realized in a microscopic model. We can illustrate it on electrons hopping on a simple square lattice (Figs.~\ref{fig_FM_AM_AFM_AAM}a,c) with an onsite (Hubbard) repulsion $U$. While the crystal potential modifies the non-interacting energy dispersion $E^0(\boldsymbol{k})$, the latter has the same form as the one introduced above close enough to the $\boldsymbol\Gamma$ point. The interaction $U$ may drive a magnetic instability, whose nature depends on the model's parameters. For instance, if the dispersion is such that the density of states has a strong peak at the Fermi level (e.g., a van Hove singularity), a ferromagnetic state may emerge through the Stoner mechanism for small enough $U$, leading to the $l=0$ spin-channel Pomeranchuk instability. On the other hand, a weak-coupling instability towards an antiferromagnetic N\'eel state is possible if the dispersion has nesting features. The antiferromagnetic state also emerges in the strong-coupling regime at half filling inside the Mott insulating state. Thus, an $l=2$ spin-channel Pomeranchuk instability likely requires other types of interactions. 

The altermagnetic ordering mechanism is fundamentally different, as here the crystal potential plays an essential role. In Sec.~\ref{AM}, we have illustrated this on the model square Lieb lattice (Figs.~\ref{fig_FM_AM_AFM_AAM}b,\ref{fig_AM}). In contrast to the simple square lattice analyzed above, the next-nearest-neighbor hopping parameters are different along the $x$ and $y$ directions, manifested in a hopping anisotropy $t_{\mathrm{a}}$ \cite{Antonenko2024}. The onsite Hubbard repulsion $U$ can still drive an instability towards an antiparallel alignment of the spins\cite{Roig2024}. Expanding the electronic dispersion around the $\boldsymbol\Gamma$ point, $E_\sigma(\boldsymbol{k}) = E^0(\boldsymbol{k}) + \frac{N t_{\mathrm{a}}}{4t} \, (k_x^2 - k_y^2)\sigma$, where $N$ is the exchange energy scale and $t$ is the nearest-neighbor hopping. The dispersion has the same form as the electronic dispersion of the ordered state generated by the $l=2$ spin-channel Pomeranchuk instability.  However, in the altermagntic ordering, the prefactor of the second term is a product $N t_{\mathrm{a}}$, instead of the single parameter $\phi$ in the Pomeranchuk instability case\cite{Roig2024}. This illustrates that both the effects of the single-particle crystal potential (encoded here in  $t_{\mathrm{a}}$) and of the many-body interactions (encoded in $N$) are required to yield an altermagnetic state. In contrast, in the case of the Pomeranchuk instability, there is a single and purely electronic energy scale (encoded in  $\phi$). 

Importantly, for a given crystal potential, there are different known routes by which interactions can drive an instability towards an antiparallel configuration of spins. For instance, besides the weak-coupling scenario outlined here, a strong-coupling Mott insulating state could also host the same spin configuration. This contributes to the much larger versatility of altermagnetism as compared to the Pomeranchuk-instability scenario, since the former can be realized in metals, insulators, and semiconductors, whereas the latter is only realized in metals.

Finally, we point out that the more detailed theoretical exploration of the spin-channel Pomeranchuk instabilities was largely motivated by spintronics\cite{Wu2004}, a pattern that was repeated fifteen years later in altermagnets\cite{Smejkal2020,Smejkal2022AHEReview,Smejkal2022a}. Also here, however, there is a significant difference. A detailed exploration of the spin-channel $l=1$ $\beta$-phase Pomeranchuk instability followed shortly after the theoretical predictions of the spin-Hall charge-to-spin conversion phenomenon in non-magnetic systems, generated by the relativistic spin-orbit coupling\cite{Murakami2003,Sinova2004}.  The time-reversal invariant spin texture in the   $l=1$ $\beta$-phase instability was called a dynamically generated spin-orbit coupling because of the resemblance  to the relativistic spin texture\cite{Wu2004}. It was emphasized that compared to the perturbatively weak  relativistic spin-orbit coupling, proportional to $1/c^2$ where $c$ is the speed of light, the dynamically generated spin-orbit coupling could lead to a significantly larger charge-to-spin conversion efficiency. However, an experimental realization of the   spin-channel $l=1$ $\beta$-phase (or $\alpha$-phase) Pomeranchuk instability has remained elusive.

The search for altermagnets was motivated by a different spintronics incentive. It followed after several years of intense research of spintronics based on collinear antiferromagnets with spin-degenerate band structures\cite{Jungwirth2018}. The driving idea was to leverage the superior spatial, temporal and energy scalability demonstrated in antiferromagnetic spintronic devices, stemming from the compensated magnetic ordering, while having well separated and conserved spin-up and spin-down channels that underpin the technologically successful spintronics based on ferromagnets\cite{Smejkal2020,Smejkal2022AHEReview,Smejkal2022a}.  As reviewed in Refs.~\onlinecite{Smejkal2022AHEReview,Smejkal2022a}, such a combination of merits, traditionally considered as mutually exclusive, is enabled by the altermagnetic ordering. Besides spintronics, the extraordinary nature of altermagnetism is projected to be favorable in a range of research fields from topological magnetism to hybrid systems integrating altermagnetism with semiconducting, superconducting, or  ferroelectric phases, as reviewed in Refs.~\onlinecite{Smejkal2022AHEReview,Smejkal2022a,Jungwirth2024}.

%In summary, altermangets share the spin-ordered anisotropic zero-magnetization nature with the A-phase of superfluid $^3$He praised for its sophistication by Anthony Leggett in the Nobel lecture quote we used to open this article.  In the case of altermagnets, however, the order is robust in many systems at room temperature and may end up being practically useful.  

%**************************************************************************
\subsection{Spin arrangements on crystals beyond altermagnetism}
\label{beyond}
In  the previous section, we have left aside an apparent conflict between the theoretically considered  ordered phases generated by the $l=1$ Pomeranchuk instabilities, and the theorem by Bloch mentioned in Sec~\ref{spin-independent}, which states that the minimum energy state bears no current\cite{Hoddeson1987}.  The diverging susceptibilities at critical values of  the Landau Fermi-liquid parameters, $F^{\rm c(s)}_l=-1$, were identified by Pomeranchuk in all angular-momentum channels including $l=1$. This  seems to violate the above theorem. Before discussing spin arrangements on crystals beyond the even-parity-wave altermagnets, in which $p$-wave magnetism has been recently predicted\cite{Hellenes2023,Hellenes2023a},  it is thus desirable to first revisit this apparent conflict in the theory of Fermi-liquid instabilities. 

The divergences of the charge and spin-channel susceptibilities for $l=1$ order parameters were analyzed in Refs.~\onlinecite{Kiselev2017,Wu2018,Klein2019}. The studies started from the exact expression for the static susceptibility for a generic order parameter. It goes beyond the Pomeranchuk $\sim 1/(1+F^{\rm c(s)}_l)$ form by including quasiparticle states away from the Fermi surface\cite{Leggett1965}. For $l=1$ order parameters corresponding to the current of conserved charge or spin, the divergencies in the generalized susceptibilities indeed disappear, regardless of the quasiparticle dispersion, i.e., both for the free-space Fermi liquid and in the presence of a crystal lattice\cite{Kiselev2017,Wu2018}. This is consistent with the above theorem by Bloch. 

However, two possibilities were identified that could enable the ordering phase transitions by the charge or spin-channel $l=1$  Pomeranchuk instabilities. One is a generic form of the $l=1$ order parameter which does not correspond to the charge or spin current\cite{Wu2018}. The other one follows from the observation that even in the cases where the static susceptibility is non-diverging for the charge or spin-current order parameter, the instability of the Fermi-liquid ground state can still be signalled by the dynamic susceptibility\cite{Klein2019}. In conclusion, present theories do not exclude the ordered $l=1$ phases, even within the framework of the momentum-space Fermi-liquid instabilities. Their real physical realizations  have, however, remained elusive.

This brings us to the recently predicted material realizations of odd-parity-wave magnetism (Fig.~\ref{fig_FM_AM_AFM_AAM}d)\cite{Hellenes2023,Hellenes2023a}. The characteristic symmetry of the corresponding spin arrangements on crystals  is time reversal combined with translation, which can be realized by an antiferroic ordering of local atomic dipoles in the crystal lattice. This symmetry is absent in altermagnetism, while it is reminiscent of conventional collinear N\'eel antiferromagnetism (Fig.~\ref{fig_FM_AM_AFM_AAM}c). 
%We thus label the phase in Figs.~\ref{FM_AM_AFM_AAM}d  and \ref{fig_He_PI_AM} "antialtermagnetic". 
However, in addition to the symmetry combining time reversal with translation, the spin arrangement on the crystal lattice shown in Fig.~\ref{fig_FM_AM_AFM_AAM}d is non-collinear coplanar, and spontaneously breaks parity symmetry. The non-collinearity can originate from, e.g.,  frustrated exchange interactions on the crystal lattice even in the absence of spin-orbit coupling, or from Dzyaloshinskii-Moriya interaction. (Even in the case where spin-orbit coupling contributes to the stabilization of the non-collinear spin ordering, the spin-ordering symmetry can be described by spin groups.) As a result, on one hand, the electronic structure in the momentum-space has the time-reversal symmetry, like in the collinear N\'eel antiferromagnetism, since a position-space translation does not change the momentum. On the other hand, however, it can show a nodal higher-partial-wave ordering in the momentum space with an alternating sign of a collinear spin polarization, and with broken both the spin-space and the real-space rotation symmetries, like in altermagnetism. Hence we label this odd-parity phase ``antialtermagnetic", and show a $p$-wave example \cite{Hellenes2023} in Fig.~\ref{fig_FM_AM_AFM_AAM}d. In the momentum space, it has analogous symmetries to the $\alpha$-phase $l=1$ Pomeranchuk instability (Figs.~\ref{fig_He_PI_AM}b,c).

The momentum-space spin-textures, analogous to the $\beta$-phase $l=1$ Pomeranchuk instability, can be realized on crystals with a non-coplanar instead of  coplanar  spin arrangement, while keeping the symmetry combining time reversal with translation, and the broken parity symmetry\cite{Hellenes2023a}. This is shown in Figs.~\ref{fig_He_PI_AM}e,f. Finally, in Fig.~\ref{fig_He_PI_AM}f we also illustrate an example of non-collinear coplanar spin arrangements on crystals without the symmetry combining time reversal and translation\cite{Smejkal2022AHEReview,Chen2014,Kubler2014,Nakatsuji2015,Nayak2016}, and with a symmetry combining spin-space and real-space rotations, resulting in a momentum-space spin texture analogous to the $\beta$-phase $l=2$ Pomeranchuk instability (Fig.~\ref{fig_He_PI_AM}e).

%**************************************************************************
\subsection{Summary}
\label{summary}
In this Perspective we have discussed the unconventional spin-ordering of the altermagnetic phase. The extraordinary nature of altermagnetism is that it spontaneously breaks the continuous spin-space rotation symmetry and the discrete real-space rotation symmetry of the crystallographic point group, while preserving a symmetry combining rotation transformations in the spin space and the real space. We have discussed key distinctions in symmetry and in microscopic ordering mechanisms between altermagnetism and ordered phases generated by  momentum-space instabilities of a Fermi liquid to shed light on the abundance and robustness of altermagnetism. This can be summaized in the following points:

%\begin{table}[h!]
%	\centering
%	\includegraphics[width=1\linewidth]{Fig5}
%	\caption{
%	\textbf{
%	Summary of analogies and distinctions between the A-phase of superfluid $^3$He and altermagnets.
%}
		% 
%First five rows highlight the analogies while the last three rows highlight the distinctions.
%%While symmetry-wise the two phases are analogous, we highlight the distinctive supporting interaction stabilizing the ordered phase which in case of superfluid $^3$He is due to subtle magnetic fluctuations while in altermagnets it is the robust crystal potential. 
%}
%\label{tab}
%\end{table}

(i) The A-phase of superfluid $^3$He was an experimental surprise and its stability is limited to a narrow range of low temperatures and high pressures. Material realizations of the $\alpha$-phase Fermi-liquid instabilities (as well as other spin-channel $l>0$ Pomeranchuk instabilities) have remained elusive. In contrast, altermagnetism was predicted by the systematic spin-symmetry group classification in a large family of materials and confirmed by microscopic density-functional-theory calculations.  The theory has guided the initial experimental verifications in room-temperature altermagnets. 

(ii) The narrow stability range of the A-phase of superfluid $^3$He was ascribed to the interplay of internal interactions in the $^3$He Fermi fluid, namely the short-range repulsion, the long-range attractive van der Waals interaction, and the effective pairing-dependent interaction due to spin fluctuations. Altermagnetism is distinct in that the ordering is stabilized by the internal electron-electron (exchange) interaction, together with the external single-particle potential of the static crystal-lattice. In altermagnetism, the crystal lattice plays an inherent part in the robust microscopic ordering mechanism. In contrast, $^3$He is a homogenous fluid with no underlying crystal lattice. In the theories of  Pomeranchuk Fermi-liquid instabilities, the crystal potential, if considered, enters only indirectly via a modified single-particle energy dispersion in the momentum-space.

(iv) Ordered phases generated by the higher-partial-wave Pomeranchuk instabilities of a Fermi-liquid are predicted to require interaction strength exceeding the critical value for the $l>0$ spin-channel instability, while avoiding the conventional $l=0$ ferromagnetic instability. In altermagnetism,  the microscopic ordering mechanism is effective from weak to strong interaction regimes, and a ferromagnetic component is excluded by the spin-group symmetry.

As a result of these distinctive features, altermagnetism is abundant and robust which suggests that its unconventional ordering can open fruitful research directions in both science and technology. 
%\begin{table} 
%\begin{tabular}{|c|c|c|}
%\hline & Superfluid ${ }^3 \mathrm{He}-\mathrm{A}$ & Altermagnets \\ \hline
%\hline 1. Broken spin rotation & Yes & Yes \\
%\hline 2. Broken orbital/lattice rotation & Yes ( $L=1$ ) & Yes $(I=2,4,6)$ \\
%\hline 3. Broken time-reversal & Possible & Yes \\
%\hline Magnetization & Vanishing & Vanishing \\
%\hline Symmetry theory neglects & Dipolar interaction among nuclear spins & Spin-orbit coupling \\
%\hline Symmetry theory & Legget, RMP (1975) & PRX 12, 031042 (2022) \\
%\hline Spin character & Composite spin of Cooper pairs ( $\mathrm{S}=1$ ) & Electron spin ( $s=1 / 2)$ \\
%\hline Realization & @ 30 bar and 2 mK & MnTe and CrSb @ ambient conditions and $>300$ material candidates \\
%\hline
%\end{tabular}
%\end{table} 

%\begin{figure}[h!]
%	\centering
%	\includegraphics[width=1\linewidth]{Fig_SC-M}
%	\caption{
%	\textbf{}
		% 
%
%}
%\label{fig:SC-M}
%\end{figure}
%\newpage

\subsection*{Acknowledgments}
We acknowledge helpful comments and suggestions on the manuscript by Anthony Leggett. TJ acknowledges support by the Ministry of Education of the Czech Republic CZ.02.01.01/00/22008/0004594 and, ERC Advanced Grant no. 101095925, RMF by the Air Force Office of Scientific Research under Award No. FA9550-21-1-0423, EF by the US National Science Foundation grant DMR 2225920 at the University of Illinois, AHM  by the Robert A. Welch Foundation under Grant Welch F-2112 and by the Simons Foundation., and JS and LŠ acknowledge support by Deutsche Forschungsgemeinschaft (DFG, German Research Foundation) - DFG (Project 452301518) and TRR 288 – 422213477 (project A09). LŠ acknowledges support from the ERC Starting Grant No. 101165122.

%\bibliographystyle{naturemag}  % ama, nar, alpha, plain, chicago, abbrv, siam
%\bibliography{Refs}

\end{document}